\def\year{2022}\relax
\documentclass[letterpaper]{article} 
\usepackage{aaai22}  
\usepackage{times}  
\usepackage{helvet}  
\usepackage{courier}  
\usepackage[hyphens]{url}  
\usepackage{graphicx} 
\usepackage{natbib}  
\usepackage{caption} 
\DeclareCaptionStyle{ruled}{labelfont=normalfont,labelsep=colon,strut=off} 
\usepackage{algorithm}
\usepackage{algorithmic}

\usepackage{afterpage}
\usepackage{booktabs}
\usepackage{multirow}
\usepackage[normalem]{ulem}
\newcommand\hl{\bgroup\markoverwith
  {\textcolor{yellow}{\rule[-.5ex]{.1pt}{2.5ex}}}\ULon}
  \usepackage[acronym]{glossaries}
\usepackage{color}
\usepackage{newfloat}
\usepackage{listings}
\floatstyle{ruled}
\newfloat{listing}{tb}{lst}{}
\floatname{listing}{Listing}
\title{Gender Gaps in Online Social Connectivity, Promotion and Relocation Reports on LinkedIn}

\author{
    Ghazal Kalhor, \textsuperscript{\rm 1}
    Hannah Gardner, \textsuperscript{\rm 2}
    Ingmar Weber, \textsuperscript{\rm 3}
    Ridhi Kashyap \textsuperscript{\rm 2}
}
\affiliations{
    \textsuperscript{\rm 1}University of Tehran, Iran\\
    \textsuperscript{\rm 2}University of Oxford, UK\\
    \textsuperscript{\rm 3}Saarland University, Germany\\
    kalhor.ghazal@ut.ac.ir, hannah.gardner@wadham.ox.ac.uk, iweber@cs.uni-saarland.de, ridhi.kashyap@nuffield.ox.ac.uk

    \vspace{0.5cm}
    \textbf{Accepted and forthcoming at the International AAAI Conference on Web and Social Media (ICWSM) 2024.}\\
    \vspace{0.5cm}
}

\usepackage{bibentry}

\begin{document}

\maketitle
\let\thefootnote\relax\footnotetext{\scriptsize This work is licensed under a Creative Commons Attribution 4.0 International License (CC BY 4.0).}

\begin{abstract}
Online professional social networking platforms provide opportunities to expand networks strategically for job opportunities and career advancement. A large body of research shows that women’s offline networks are less advantageous than men’s. How online platforms such as LinkedIn may reflect or reproduce gendered networking behaviours, or how online social connectivity may affect outcomes differentially by gender is not well understood. This paper analyses aggregate, anonymised data from almost 10 million LinkedIn users in the UK and US information technology (IT) sector collected from the site’s advertising platform to explore how being connected to Big Tech companies (‘social connectivity’) varies by gender, and how gender, age, seniority and social connectivity shape the propensity to report job promotions or relocations. Consistent with previous studies,  we find there are fewer women compared to men on LinkedIn in IT. Furthermore, female users are less likely to be connected to Big Tech companies than men. However, when we further analyse recent promotion or relocation reports, we find women are more likely than men to have reported a recent promotion at work, suggesting high-achieving women may be self-selecting onto LinkedIn. Even among this positively selected group, though, we find men are more likely to report a recent relocation. Social connectivity emerges as a significant predictor of promotion and relocation reports, with an interaction effect between gender and social connectivity indicating the payoffs to social connectivity for promotion and relocation reports are larger for women. This suggests that online networking has the potential for larger impacts for women, who experience greater disadvantage in traditional networking contexts, and calls for further research to understand  differential impacts of online networking for socially disadvantaged groups.
\end{abstract}

\section{Introduction}
Maintaining effective professional networks is important for career progression. Within a broader landscape of the digitalisation of work, online professional social networking sites (SNSs) offer widespread opportunities for workers to manage their image and career, connect with professional networks, and find new job roles. With over 875 million users, LinkedIn is the world's largest professional networking platform \footnote{As of Jan 14, 2023: \url{https://about.linkedin.com}.}. Although the large user base engaging with LinkedIn as part of their career strategy makes it an important platform to study in its own right, recent work also shows that online networking behaviour on LinkedIn influences job opportunities for users \cite{rajkumar2022causal, wheeler2022linkedin}.

Given the potentially significant benefits of online professional networking, it is important to understand how different users, especially those who have conventionally faced greater disadvantages in the labour market, such as women, engage with online professional networks. A large body of research shows that women's offline networks are less advantageous than men's \cite{Wang2009, mengel2020gender}. How online platforms such as LinkedIn may reflect or reproduce gendered networking behaviours \cite{Bapna2018, Greguletz2018, Wang2009, Kirton2018}, or how online social connectivity may affect outcomes differentially by gender is not well understood. Although existing research highlights gender gaps in LinkedIn use \cite{Kashyap2021}, with larger gaps in certain industries such as information technology (IT) \cite{Haranko2018ProfessionalGG, verkroost2020tracking}, other dimensions such as gender gaps in social connectivity and its link with job progression outcomes, such as promotion or relocation, have received limited attention. This paper examines these dimensions using aggregated, anonymised data from almost 10 million LinkedIn users from the site's advertising platform. Specifically, we examine how gender, age, seniority, and connectedness to big companies (what we term social connectivity) are associated with the propensity to report job promotions or relocations on the LinkedIn population.

Our study focuses on LinkedIn users within the IT sector in the US and UK, a male-dominated set of industries within which progression for women can be made particularly difficult through pervasive gender stereotypes about technical skills, `geek' culture, `old boys' clubs of informal networks, and organisational factors such as long hours that increase the difficulty to achieve work-family balance \cite{Ahuja2002, Armstrong2018}. Socialising and networking events within IT companies are commonly reported to be in male-oriented spaces -- e.g., involving sports, pub trips, and take place outside of already long industry working hours \cite{Cross2006, Kirton2018, McGee2018, earles2020gender} -- with some women feeling uncomfortable or unable to attend as often as male colleagues due to gendered family expectations, and others not receiving invitations \cite{Bjerk2008, Kirton2018}. These norms within the industry may further reinforce disadvantages faced by women within it, and limit their ability to expand their networks in beneficial ways for career progression. In contrast, online networking theoretically offers greater flexibility to participate in terms of time and location, with the potential to bolster opportunities for those that face greater constraints, such as women. It may provide opportunities to expand and build advantageous networks beyond those encountered within one's immediate work environment. However, whether online connectivity provides these differential payoffs by gender has received limited empirical attention, and is a question that our study examines.

\section{Related Work}
Gender differences in building, maintaining, and using offline professional networks have been widely investigated across different settings. Women’s networks are generally characterised as smaller groups formed of stronger ties, and more intimate connections than men’s \cite{Wang2009, Greguletz2018}. Within the IT sector, leveraging a natural experiment, Bapna and Funk find gender differences in network formation, with women at an IT conference meeting 42\% fewer new contacts, spending 48\% less time talking with them, and adding 25\% fewer LinkedIn connections than men \cite{Bapna2018}. Mengel proposes that it is largely differences in network use across genders that leads to higher payoffs for men \cite{mengel2020gender}. Their lab experiments indicate that men reward their network neighbours more than women do, and as men’s networks show greater gender homophily, these benefits are disproportionately passed onto other men. Presenting yourself to potential networking contacts, interviewing successfully, and negotiating promotions all aid career advancement, but require a person to promote themselves to others. In professional contexts, research has shown that women consistently underperform when asked to self-promote, both in self-assessment and when judged anonymously by others \cite{MossRacusin2010, Smith2013, Rudman1998}. The gendered expectation that women will act modestly has been shown to drive this underperformance \cite{Rudman1998}, with women altering behaviour to avoid backlash from others who may derogate them for bragging \cite{Lindeman2018}. These studies suggest that women's career progression is disfavoured in traditional network-based professional contexts. 

However, how these differences in networking and professional self-promotion behaviours have been translated, reproduced or offset within the context of new online professional spaces is not clear. LinkedIn, as the world’s largest professional networking platform, is an important context in which to understand how men and women use online spaces differentially, because recent experimental work has shown that LinkedIn use has real job market implications for users. Rajkumar and colleagues suggest there is a causal relationship between LinkedIn's creation of new ties in networks and their translation into job opportunities \cite{rajkumar2022causal}, particularly highlighting the importance of moderately weak ties over strong ties in job transmission. Wheeler et al.\ suggest links between training job-seekers to use LinkedIn, and increased employment rates \cite{wheeler2022linkedin}. Both suggest that connectivity within online networks brings important benefits to users in their working lives, although these studies do not examine heterogeneity of these effects by gender. 

A body of work is beginning to describe gender gaps found across LinkedIn, showing how the platform is a gendered space for interaction, with differences in how men and women use their profiles to self-present. Altenburger et al. find that female MBA graduates are less likely to use free-form data fields on their profiles (such as the Summary and Job Description fields) but are just as likely to include structured fields (such as Honours and Skills) \cite{altenburger2017there}. Similarly, another study suggested that men were more likely to receive and give recommendations, and to display personal and professional interests, but were unable to control for industry confounding in these behaviours \cite{Zide2014}. Aguado later drew on this work but found conversely that women and more senior individuals showed greater breadth of interaction on LinkedIn (encompassing recommendations given, companies and people followed, and skills validated) \cite{Aguado2019}. Women were also more likely to have completed additional sections of their profiles (such as interests identified, length of written text, languages identified). However, less is known about how these online gender differentials, particularly in connectivity to potentially advantageous users, are related to professional outcomes. 

To the best of our knowledge, only four previous studies have harnessed data from LinkedIn’s advertising platform as we do to study gender gaps on LinkedIn. Haranko et al.\ analyse gender gaps on LinkedIn across 20 US cities and find that there is little variation across location, but larger variation exists across industries, with the high-tech industry as among the most gender imbalanced \cite{Haranko2018ProfessionalGG}. The authors also find technical, and computing skills reported on LinkedIn to be highly male-dominated. \citet{Berte2023MonitoringGG} examine subnational variations in gender gaps on LinkedIn in Italy, and consistent with regional labour market gender gaps find that women are also underrepresented on LinkedIn. Kashyap and Verkroost analyse gender gaps on the platform across countries, ages, industries and seniorities, and find women to be significantly underrepresented on LinkedIn in Science, Engineering, Maths and Technology (STEM) fields, as well as in higher-level managerial positions \cite{Kashyap2021}. Our work builds on these aforementioned studies, as well as \citet{verkroost2020tracking}, which computes gender gap indices (GGIs) using LinkedIn’s advertising platform data, describing variation across countries and industries within the IT sector specifically. We extend these studies by looking at gender differences in LinkedIn across additional characteristics and behaviours, such as social connectivity, promotion, and relocation reports. 

\section{Data}
Part of LinkedIn’s revenue is generated by offering an advertising platform, allowing advertisers to reach over 875 million LinkedIn global users\footnote{\url{https://about.linkedin.com/}}. To maximise the effectiveness of advertising campaigns, LinkedIn offers advertisers the possibility to carefully target their audience based on a number of user attributes. The available attributes are based on a combination of self-declared information, and information inferred using machine learning from user activity and user profiles. For example, users' employment history or social network connections are based on the information explicitly provided by them in their profiles. On the other hand, age and gender are inferred from profile information, including ``the pronouns used when others recommend [them] for skills\footnote{\url{https://www.linkedin.com/help/linkedin/answer/a517610/inferred-age-or-gender-on-linkedin}}. Similarly, information on the user’s job seniority is inferred, most likely based on job titles. Information about a user's location is likely based on a combination of self-provided employment history and IP-based geo-location\footnote{\url{https://www.linkedin.com/help/lms/answer/a422631}}. Advertisers can then target their advertisements to users with a desired combination of attributes.

To launch and manage advertising campaigns, LinkedIn provides advertisers with an online platform\footnote{\url{https://www.linkedin.com/campaignmanager}}. As part of the campaign and budget planning process, the advertising platform provides advertisers with so-called “audience estimates”, estimating how many LinkedIn users match the provided targeting criteria selected by the advertiser. These aggregated estimates, which are provided free of cost before launching an advertising campaign, create a kind of digital census: for any chosen set of targeting attributes, potential advertisers can obtain a count estimate of how many of its users match the targeting criteria. These estimates can be collected programmatically through an API\footnote{\url{ https://worldbank.github.io/connectivity_mapping/linkedin_nbs/interface.html}}.

For our analysis, we collected a large number of such individual audience estimates by repeatedly modifying the targeting criteria provided to the advertising platform. We decided to limit our data collection to cover LinkedIn users in the US and the UK, two of the largest user populations on LinkedIn \cite{Kashyap2021} that are also culturally similar. We further narrowed our data population by only collecting audience estimates for LinkedIn users currently employed in the information technology (IT) sector, which we define by the company industry of users, covering the following 11 industries: Internet, Information Technology and Services, Computer Software, Computer and Network Security, Computer Hardware, Computer Networking, Wireless, Telecommunications, Semiconductors, Nanotechnology, and Consumer Electronics. Aligned with previous work \cite{verkroost2020tracking}, our definition of the IT sector relies on the OECD definition of the ICT industry \cite{wunsch2007directorate}, as defined according to International Standard Industry Classification (ISIC) Revision 4. This definition includes manufacturing-related industries, such as electronics and semiconductors, in addition to service-related parts of the industry, like software and internet services. It is worth noting that the names of some of these industries on LinkedIn have been changed from 2021 to 2023. Therefore, we have provided the mapping of their previous names to their current names in Table \ref{tab:tab1}.
For the above-mentioned selection, we then looped over combinations of (i) a user’s location, either the US or the UK, (ii) their inferred job seniority, (iii) whether they recently reported a promotion or relocation, (iv) their gender, (v) their age range (vi), and whether they have a social network connection to an employee at (at least) one of several big companies, namely, Facebook, Apple, Amazon Web Services, Microsoft, and Google. Together, these companies are often referred to as the Big Five and are seen as the quintessential representatives of Big Tech \cite{Birch2022}. As many connections to employees at these companies are likely to come from colleagues at the same companies, we excluded LinkedIn users who were at the time of data collection working at any of the Big Five companies. Using this measure of social connectivity captures the influence of external social networks at prestigious companies. Prestigious external networks can have impacts on job prospects of an individual through serving as better information access for the labour market, through exposure to new ideas from established companies, and the potential for external job offers from these companies. Together, each of these can increase perceived desirability of employees,  which may make an individual's existing company more likely to retain or promote them, or make them competitive in the job market more generally. 

\begin{table*}
\centering
  \caption{Mapping of previous names of industries to current names.}
  \label{tab:tab1}
  \begin{tabular}{ccl}
    \toprule
    Previous Name & Current Name\\
    \midrule
    Internet & Technology, Information and Internet\\
    Information Technology and Services & IT Services and IT Consulting\\
    Computer Software & Software Development\\
    Computer Hardware & Computer Hardware Manufacturing\\
    Computer Networking & Computer Networking Products\\
    Wireless & Wireless Services\\
    Semiconductors & Semiconductor Manufacturing\\
    Nanotechnology & Nanotechnology Research\\
    Consumer Electronics & Computers and Electronics Manufacturing\\
  \bottomrule
\end{tabular}
\end{table*}

Users are considered to have relocated if they have ``recently relocated their permanent location", while LinkedIn infers recent promotions based on user-provided profile updates to employment history. Table \ref{tab:tab2} provides an overview of the features that we collected. This data was collected in June and July of 2021. Data were preprocessed and stored as a CSV file. We analysed the counts of users for each combination of these variables (total of 192 unique combinations, 2 (gender) $\times$ 4 (seniority) $\times$ 4 (age range) $\times$ 3 (recent status) $\times$ 2 (social connectivity)) to make sure they are not zero or, in other words, to remove sparsity from the dataset. Our final dataset consisted of 156 combinations of the variables, as 36 were dropped due to small audience counts. More specifically, to protect user privacy, LinkedIn does not provide audience estimates when the targeted audience is smaller than 300.
Note that, apart from sparsity, our aggregate data is conceptually \emph{equivalent} to individual-level data for the set of covariates considered, as our data contains all possible cross-tabulations. The dataset and analysis code supporting the conclusions are available at \url{https://github.com/kalhorghazal/icwsm-promotionrelocation-gendergaps}.

\begin{table}
\centering
  \caption{The dataset features and their possible values.}
  \label{tab:tab2}
  \begin{tabular}{ccl}
    \toprule
    Feature&Possible Values\\
    \midrule
    Gender & Female, Male\\
    Job Seniority & Entry, Senior, Manager, Director\\
    Age Range & 18 to 24, 25 to 34, 35 to 54, 55+\\
    Recent Status & Promoted, Relocated, Any\\
    Social Connectivity & Connected to big companies, Any\\
    Count & Integer Values $\geq 300$\\
  \bottomrule
\end{tabular}
\end{table}

\section{Results}
\subsection{Gender Differences in LinkedIn Use}
Our data show that there are fewer women than men in the IT sector on LinkedIn, consistent with previous work on LinkedIn \cite{verkroost2020tracking}. While this gender gap on LinkedIn may reflect there are fewer women working in the IT sector \cite{McKinney2008}, it may also reflect how women working within the IT sector select into being LinkedIn users. The distribution across age groups differs between men and women, with Figure \ref{fig:fig3} showing that women on the platform and working in the IT sector are proportionately younger than men. Women aged 25 to 34 make up half of the female professional population in IT on LinkedIn, while the male distribution is flatter, with similar proportions of workers in both the 25 to 34 and 35 to 54 age categories. Lower proportions of women older than 35 perhaps reflect workforce departure after family formation, or increasing gender balance in entry to the sector over time. We also find that women in IT on LinkedIn are more junior than their male counterparts. Figure \ref{fig:fig3} shows distributions of women and men across seniority levels to be largely similar at the senior and manager levels, but that a higher proportion of all women are working in entry-level positions than men, and a correspondingly lower proportion of women are employed at the highest director rank. The results of a Kolmogorov-Smirnov test \cite{hodges1958significance} indicate these differences between women's and men's job seniority distributions is highly statistically significant ($p< 10^{-16}$).

Our next analyses of the promotion and relocation reporting behaviours hence normalises the male and female populations by age and seniority distribution to make it possible to isolate differences attributable to gender using demographic standardisation methods \cite{klein2001age}. When applying adjustment to our calculations, we consider gender-agnostic age or seniority distributions on LinkedIn as the reference distribution.
\afterpage{
\begin{figure*}[t!]
  \centering
  \includegraphics[width=\linewidth]{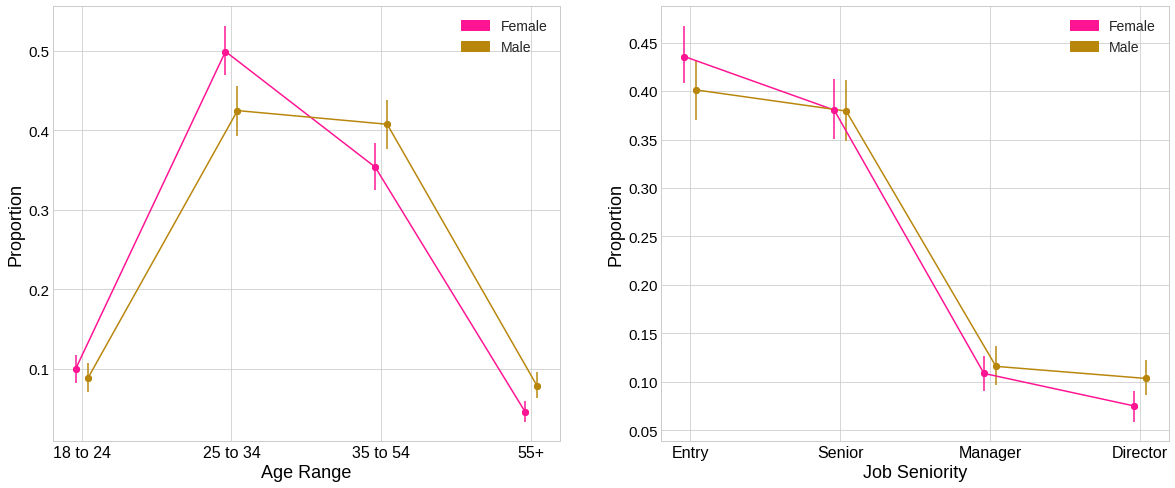}
  \caption{Age range and job seniority distributions of users disaggregated by gender. 95\% confidence intervals shown, with standard errors computed via bootstrapping.}
  \label{fig:fig3}
\end{figure*}

\begin{table*}[h!]
\centering
  \caption{Age and seniority adjustment calculations for promotion report ratios by gender.}
  \label{tab:tab3}
  \begin{tabular}{cccccccccl}
    \toprule
    \begin{tabular}{@{}c@{}}Age Range / \\ Seniority Level\end{tabular} &
    \multicolumn{2}{c}{\begin{tabular}{@{}c@{}}Number of Promotions \\ $(a)$\end{tabular}} & 
    \multicolumn{2}{c}{\begin{tabular}{@{}c@{}}Population (Millions) \\ $(b)$\end{tabular}} & 
    \multicolumn{2}{c}{\begin{tabular}{@{}c@{}}Rate per 100,000 \\ $(c=(a / b) \times 100,000)$\end{tabular}} & 
    \multirow{2}{*}{Weight $(d)$} &
    \multicolumn{2}{c}{\begin{tabular}{@{}c@{}}Weighted Rate \\$(c \times d)$\end{tabular}} \\
    \midrule
     & 
    Female & Male & 
    Female & Male &
    Female & Male &
    & 
    Female & Male \\
    18-24 & 
    5880 & 8000 & 
    0.36 & 0.54 &
    1633.3 & 1481.5 &
    0.093 & 

    151.9 & 137.8 \\
    25-34 & 
    21390 & 30290 & 
    1.81 & 2.60 &
    1181.8 & 1165.0 &
    0.452 & 

    534.2 & 526.6 \\    
    35-54 & 
    15160 & 26600 & 
    1.28 & 2.49 &
    1184.4 & 1068.3 &
    0.388 & 

    459.5 & 414.5 \\    
    55+ & 
    690 & 1590 & 
    0.17 & 0.48 &
    405.9 & 331.2 &
    0.067 & 

    27.2 & 22.2 \\    
    Total & 
    43120 & 66480 & 
    3.62 & 6.11 &
    1191.2 & 1088.0 &    
    1.000 & 

    \textbf{1172.8} & \textbf{1101.0} \\    
    \bottomrule
    Entry & 
    1540 & 2500 & 
    1.58 & 2.45 &
    97.5 & 102.0 &
    0.414 & 

    40.4 & 42.2 \\    
    Senior & 
    21390 & 33160 & 
    1.38 & 2.32 &
    1550.0 & 1429.3 &
    0.380 & 

    589.0 & 543.1 \\    
    Manager & 
    11950 & 17260 & 
    0.39 & 0.71 &
    3064.1 & 2431.0 &
    0.113 & 

    346.2 & 274.7 \\    
    Director & 
    8240 & 13560 & 
    0.27 & 0.63 &
    3051.8 & 2152.4 &
    0.093 & 

    283.8 & 200.2 \\
    Total & 
    43120 & 66480 & 
    3.62 & 6.11 &
    1191.2 & 1088.0 &
    1.000 & 

    \textbf{1259.4} & \textbf{1060.2} \\    
    \bottomrule
  \end{tabular}
\end{table*}

\begin{table*}[h!]
\centering
  \caption{Age and seniority adjustment calculations for relocation report ratios of each gender.}
  \label{tab:tab4}
  \begin{tabular}{cccccccccl}
    \toprule 
    \begin{tabular}{@{}c@{}}Age Range / \\ Seniority Level\end{tabular} &
    \multicolumn{2}{c}{Number of Relocations} & 
    \multicolumn{2}{c}{Population (Millions)} & 
    \multicolumn{2}{c}{Rate per 100,000} & 
    Weight &
    \multicolumn{2}{c}{Weighted Rate} \\
    \midrule
     & 
    Female & Male & 
    Female & Male &
    Female & Male &
     & 
    Female & Male \\
    18-24 & 
    2640 & 5950 & 
    0.36 & 0.54 &
    733.3 & 1101.8 &
    0.093 & 

    68.2 & 102.5 \\    
    25-34 & 
    18790 & 30340 & 
    1.81 & 2.60 &
    1038.1 & 1166.9 &
    0.452 & 

    469.2 & 527.4 \\    
    35-54 & 
    17610 & 42100 & 
    1.28 & 2.49 &
    1375.8 & 1690.8 &
    0.388 & 

    533.8 & 656.0 \\    
    55+ & 
    1750 & 8100 & 
    0.17 & 0.48 &
    1029.4 & 1687.5 &
    0.067 & 

    69.0 & 113.1 \\
    Total & 
    40790 & 86490 & 
    3.62 & 6.11 &
    1126.8 & 1415.5 &
    1.000 & 

    \textbf{1140.2} & \textbf{1399.0}\\    
    \bottomrule
    Entry & 
    13490 & 27200 & 
    1.58 & 2.45 &
    1415.5 & 1110.2 &
    0.414 & 

    586.0 & 459.6 \\    
    Senior & 
    17440 & 35250 & 
    1.38 & 2.32 &
    1263.8 & 1519.4 &
    0.380 & 

    480.2 & 577.4 \\    
    Manager & 
    4760 & 8480 & 
    0.39 & 0.71 &
    1220.5 & 1194.4 &
    0.113 & 

    137.9 & 135.0 \\    
    Director & 
    5100 & 15560 & 
    0.27 & 0.63 &
    1888.9 & 2469.8 &
    0.093 & 

    175.7 & 229.7 \\    
    Total & 
    40790 & 86490 & 
    3.62 & 6.11 &
    1126.8 & 1415.5 &
    1.000 & 

    \textbf{1379.8} & \textbf{1401.7} \\    
    \bottomrule
  \end{tabular}
\end{table*}
}
\subsection{Promotion and Relocation Reports by Gender}

Table \ref{tab:tab3} displays promotion report rates by age- and seniority- groups, and by gender. The age- and seniority-adjusted rates show that women are between 6.5\% and 18.8\% more likely than men to recently report a promotion. As shown in Table \ref{tab:tab3}, within each age category the youngest LinkedIn users are most likely to report promotions, with successively older age groups each reporting promotions at lower rates. Male and female distributions track a similar trend with increasing age, but women consistently have higher rates of promotion reports than men across all age groups. After age-adjustment, per 100,000 women, 1,173 had recently been promoted compared to 1,101 per 100,000 men, indicating a highly statistically significant difference in promotion reports by gender of 72 per 100,000 ($z=10.285$, $SE=7.000 \times 10^{-5}$, $p< 0.001$).

LinkedIn users at successively higher seniority levels are more likely to report their recent promotions, except at the director level (Table \ref{tab:tab3}). This is in line with existing work showing that senior employees are more likely to be promoted than juniors \cite{mills1985seniority}, although our data cannot differentiate between actual promotion rates of users, and the rate at which users report a promotion they have received to their online network. Promotion report rates increase most between the manager and director levels, which is also the seniority group among which there is the largest difference between male and female reports. Our findings further indicate that it is only at manager level that differences in promotion reports begin to emerge strongly, with LinkedIn's entry and senior groups showing similar rates of promotion reports for both genders. Across all except entry-level users, women have a higher promotion report rate than men. After seniority adjustment, this higher rate of female promotion reports is maintained, with 1,259 per 100,000 female users and 1,060 per 100,000 male users reporting promotion. This difference, of 199 per 100,000, is highly statistically significant ($z=28.349$, $SE=7.020 \times 10^{-5}$, $p< 0.001$). Thus, overall, even adjusting for gender differences in age and seniority composition among LinkedIn users, we see women showing higher promotion reports.

Table \ref{tab:tab4} displays age- and seniority- adjusted rates of male and female relocation, according to LinkedIn's categorisations, and suggests that men are between 22.7\% and 1.6\% more likely to report relocation than women, respectively. After age-adjustment, per 100,000 women, 1,140 had recently relocated, compared to 1,399 per 100,000 men, indicating a difference of 259 per 100,000 fewer relocations for women ($z=34.430$, $SE=7.517 \times 10^{-5}$, $p< 0.001$). After seniority adjustment, differences between male and female relocation reports are smaller (-21.8 per 100,000) than after age-adjustment, but differences are still statistically significant ($z=2.831$, $SE=7.772 \times 10^{-5}$, $p< 0.01$). The increase in rates of female relocation after seniority adjustment indicates that either mobility among female-dominated entry positions is lower than average across ranks, and/or that mobility among male-dominated director positions is higher. As shown in Table \ref{tab:tab4}, entry-level women relocate at a higher rate than men, but among senior employees and directors, males relocate significantly more, in line with previous work showing that women are more mobile earlier in their careers \cite{branden2011whose}. For both genders, older users are more likely to report their recent relocations except for the 55+ age range, where women's likelihood of reporting relocation decreases to similar levels as 25 to 34 year-olds, but men's likelihood roughly stagnates. At all ages, women are significantly less likely to report relocation.

\subsection{Gender Differences in Social Connectivity }
\begin{figure}[ht]
  \centering
  \includegraphics[width=\linewidth]{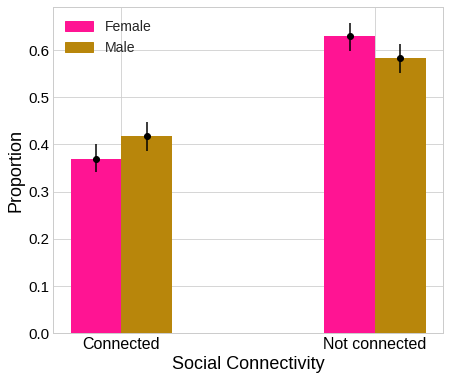}
  \caption{Bar blot of socially connected and unconnected users disaggregated by gender. 95\% confidence intervals shown, with standard errors computed via bootstrapping.}
  \label{fig:fig6}
\end{figure}

\begin{figure}[ht]
  \centering
  \includegraphics[width=\linewidth]{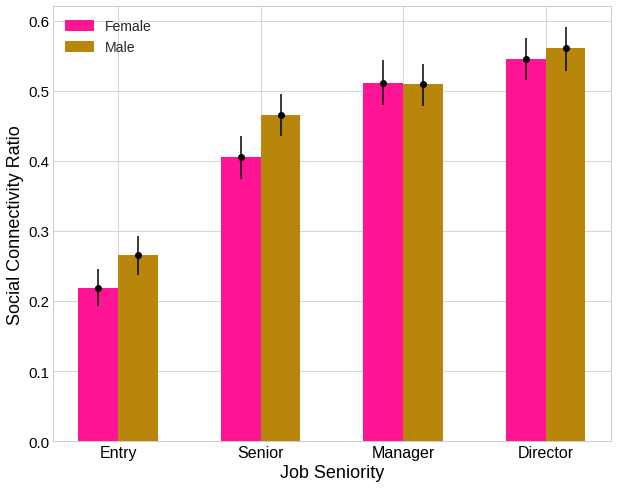}
  \caption{Gender-disaggregated bar plot of social connectivity ratio (proportion connected to big company) at each seniority level. 95\% confidence intervals shown, with standard errors computed via bootstrapping.}
  \label{fig:fig7}
\end{figure}

Women not working at a Big Tech company are less likely than men to be connected to an employee of a Big Tech company, as shown in Figure \ref{fig:fig6}, in line with literature that shows that women's networks are not as advantageous as men's \cite{greguletz2019women}. As shown in Figure \ref{fig:fig7}, social connectivity increases with seniority for both male and female users, with men having higher or the same rates of connectivity at each rank, but with smaller gender differentials at more senior Manager and Director ranks. This likely reflects a type of survivorship bias \cite{Fryer2007, Smith2013} among the women at these higher ranks -- although women are less likely to be at these higher ranks, those who are present in them are positively selected, and fairly equally socially connected to men at these ranks. 

\begin{figure}[ht]
  \centering
  \includegraphics[width=\linewidth]{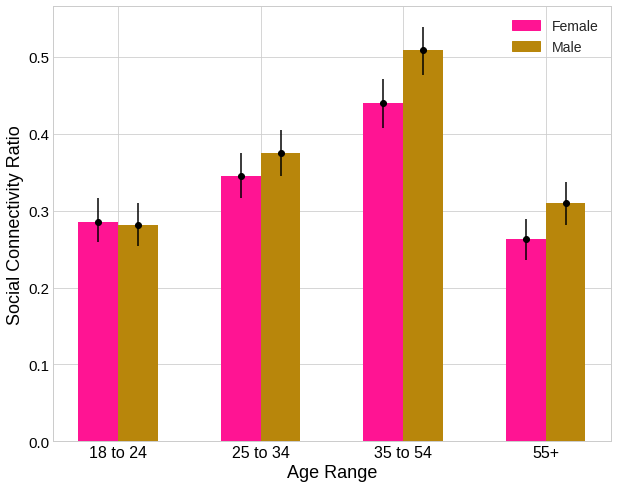}
  \caption{Gender-disaggregated bar plot of social connectivity ratio (proportion connected to big company) at each age group. 95\% confidence intervals shown, with standard errors computed via bootstrapping.}
  \label{fig:fig8}
\end{figure}

While social connectivity is similarly low for all the youngest users aged 18 to 24, a gender connectivity gap emerges and widens in the successive age groups from the ages of 25 to 54, peaking among 35--54 year olds (Figure \ref{fig:fig8}). The gap in these age groups widens across the ages at which women are often likely to experience interrupted labour market trajectories due to family formation and care-giving roles, and when professional careers are likely to become well established.  Among the oldest employees, men over 55 maintain a connectivity advantage over female colleagues, but both show social connectivity levels comparable to the youngest (18--24) age group.

\subsection{Relationship between Social Connectivity and Status Reports}
To assess the relationship between social connectivity and promotion and relocation reports for LinkedIn users, we estimate different logistic regression models \cite{kleinbaum2002logistic}, as promotion and relocation reports are dichotomous (binary) outcomes. Through this analysis, we examine whether potentially advantageous online connections are associated with job progression outcomes, and whether this association differs by gender. We estimate four models with different combinations of predictors, with social connectivity being our key predictor of interest, and compare them using Akaike Information Criterion (AIC) \cite{akaike1974new} to determine the model with the best fit (smaller AIC value). The dataset on which these regressions are estimated is 9,735,600 rows, each representing a specific user. These rows are obtained by unrolling our aggregated dataset that contains counts of each combination of targeting attributes. Table \ref{tab:tab5} shows an example of the initial aggregated dataset. Table \ref{tab:tab6} and \ref{tab:tab7} present the unrolled (individual-level) versions used for the promotion and relocation report models.

\begin{table*}
\centering
  \caption{An example of the initial dataset.}
  \label{tab:tab5}
  \begin{tabular}{ccccccl}
    \toprule
    Gender&Job Seniority&Age Range&Social Connectivity&Recent Status&Count\\
    \midrule
    Female & Senior & 35 to 54 & Connected to big companies & Any & 3\\
    Female & Senior & 35 to 54 & Connected to big companies & Relocated & 1\\
    Male & Manager & 25 to 34 & Not connected & Promoted & 3\\
  \bottomrule
  \\
\end{tabular}

\centering
  \caption{Unrolled version of the example in Table \ref{tab:tab5} for the promotion report models.}
  \label{tab:tab6}
  \begin{tabular}{cccccl}
    \toprule
    Gender&Job Seniority&Age Range&Social Connectivity&Promotion Status\\
    \midrule
    Female & Senior & 35 to 54 & Connected to big companies & Not promoted\\
    Female & Senior & 35 to 54 & Connected to big companies & Not promoted\\
    Male & Manager & 25 to 34 & Not connected & Promoted\\
    Male & Manager & 25 to 34 & Not connected & Promoted\\
    Male & Manager & 25 to 34 & Not connected & Promoted\\
  \bottomrule
  \\
\end{tabular}

\centering
  \caption{Unrolled version of the example in Table \ref{tab:tab5} for the relocation report models.}
  \label{tab:tab7}
  \begin{tabular}{cccccl}
    \toprule
    Gender&Job Seniority&Age Range&Social Connectivity&Relocation Status\\
    \midrule
    Female & Senior & 35 to 54 & Connected to big companies & Not relocated\\
    Female & Senior & 35 to 54 & Connected to big companies & Not relocated\\
    Female & Senior & 35 to 54 & Connected to big companies & Relocated\\
  \bottomrule
\end{tabular}
\end{table*}

\begin{figure*}[ht]
  \centering
  \includegraphics[width=\textwidth]{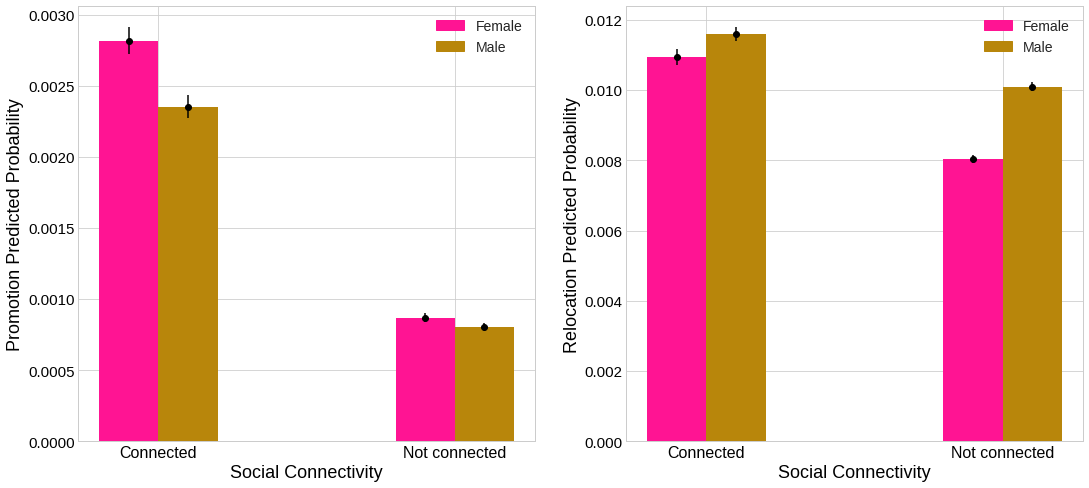}
  \caption{Predicted probabilities of the interaction between gender and social connectivity in promotion (left)/relocation (right) models. 95\% confidence intervals shown, derived from the combination of the z-scores and standard errors \cite{diez2012openintro}.}
  \label{fig:fig11}
\end{figure*}

To examine the differential impacts of social connectivity by gender, we include a social connectivity $\times$ gender interaction term in our models. We also check for potential multicollinearity of predictor variables in the regression models by calculating the variance inflation factor (VIF) of each variable. As all observed factors were smaller than 5.0 (the biggest factor was 2.8 for the interaction of gender and social connectivity), we did not exclude any predictors from the models, following standard guidelines \cite{Akinwande2015}. Table \ref{tab:tab8} shows the model estimates with the outcome of whether a user has recently reported a promotion or not. Based on the odds ratios of social connectivity, which are higher than one, we can conclude that users connected to Big Tech companies, i.e. those with potentially advantageous external networks, are more likely to report recent promotions. Consistent with our age- and seniority-specific analyses shown previously, these models also show that younger users, higher seniorities, as well as women have higher odds of reporting promotions, and that these gender differences in promotion reports persist even after we control for age and seniority. Looking at the social connectivity $\times$ gender interaction, we see a positive interaction (odds ratio higher than 1), which suggests that social connectivity has higher payoffs (stronger positive association) for women in predicting promotion. This differential impact by gender can be seen in Figure \ref{fig:fig11} that shows the predicted probabilities of promotion by gender (holding other covariates at their mean/mode values) by social connectivity from the full model shown in Table \ref{tab:tab8}. Although social connectivity boosts promotion probability for both men and women, it bolsters women's probability of reporting a promotion comparatively more than men, approximately 0.046 percent, or 46 per 100,000 promotions more, which translates to a 3.86 percent increase above the mean promotion report rate for women of 1191.2 per 100,000 (as shown in Table \ref{tab:tab3}).

\begin{table*}
\centering
  \caption{Estimates (odds ratios) from logistic regression models predicting promotion status by user characteristics.}
  \label{tab:tab8}
  \begin{tabular}{cccccl}
    \toprule
    \begin{tabular}{@{}c@{}}\textit{Dependent variable = Recently promoted} \\ \textit{(ref: Not promoted)}\end{tabular} 
    & 
    \multicolumn{4}{c}{\begin{tabular}{@{}c@{}}\textit{Odds ratio} \\ \textit{(standard error)}\end{tabular}} \\ 
    \midrule
    \begin{tabular}{@{}c@{}}Gender (ref: Male) \\ Female\end{tabular}
    & 
    \begin{tabular}{@{}c@{}}1.032** \\ (0.010)\end{tabular}
    &
    \begin{tabular}{@{}c@{}}1.142*** \\ (0.011)\end{tabular}
    &
    \begin{tabular}{@{}c@{}}1.142*** \\ (0.007)\end{tabular}
    &
    \begin{tabular}{@{}c@{}}1.079*** \\ (0.010)\end{tabular} \\

    \begin{tabular}{@{}c@{}}Job Seniority (ref: Entry) \\ Senior\end{tabular}
    & - &
    \begin{tabular}{@{}c@{}}12.572*** \\ (0.206)\end{tabular}
    &
    \begin{tabular}{@{}c@{}}14.974*** \\ (0.246)\end{tabular}
    &
    \begin{tabular}{@{}c@{}}14.964*** \\ (0.246)\end{tabular} \\

    Manager
    & - &
    \begin{tabular}{@{}c@{}}22.272*** \\ (0.376)\end{tabular}
    &
    \begin{tabular}{@{}c@{}}28.147*** \\ (0.477)\end{tabular}
    &
    \begin{tabular}{@{}c@{}}28.068*** \\ (0.476)\end{tabular} \\

    Director
    & - &
    \begin{tabular}{@{}c@{}}18.119*** \\ (0.314)\end{tabular}
    &
    \begin{tabular}{@{}c@{}}26.224*** \\ (0.459)\end{tabular}
    &
    \begin{tabular}{@{}c@{}}26.179*** \\ (0.458)\end{tabular} \\

    \begin{tabular}{@{}c@{}}Age Range (ref: 18 to 24) \\ 25 to 34\end{tabular}
    &
    \begin{tabular}{@{}c@{}}0.709*** \\ (0.007)\end{tabular}
    & - &
    \begin{tabular}{@{}c@{}}0.567*** \\ (0.006)\end{tabular}
    &
    \begin{tabular}{@{}c@{}}0.567*** \\ (0.006)\end{tabular} \\

    35 to 54
    &
    \begin{tabular}{@{}c@{}}0.518*** \\ (0.005)\end{tabular}
    & - &
    \begin{tabular}{@{}c@{}}0.294*** \\ (0.003)\end{tabular}
    &
    \begin{tabular}{@{}c@{}}0.294*** \\ (0.003)\end{tabular} \\

     55+
    &
    \begin{tabular}{@{}c@{}}0.170*** \\ (0.004)\end{tabular}
    & - &
    \begin{tabular}{@{}c@{}}0.090*** \\ (0.002)\end{tabular}
    &
    \begin{tabular}{@{}c@{}}0.091*** \\ (0.002)\end{tabular} \\

    \begin{tabular}{@{}c@{}}Social Connectivity (ref: Not connected) \\ Connected to big companies\end{tabular}
    & 
    \begin{tabular}{@{}c@{}}3.811*** \\ (0.031)\end{tabular}
    &
    \begin{tabular}{@{}c@{}}2.462*** \\ (0.020)\end{tabular}
    &
    \begin{tabular}{@{}c@{}}3.045*** \\ (0.019)\end{tabular}
    &
    \begin{tabular}{@{}c@{}}2.922*** \\ (0.024)\end{tabular} \\

    Social Connectivity $\times$ Gender
    & 
    \begin{tabular}{@{}c@{}}1.227*** \\ (0.015)\end{tabular}
    &
    \begin{tabular}{@{}c@{}}1.180*** \\ (0.015)\end{tabular}
    & - &
    \begin{tabular}{@{}c@{}}1.109*** \\ (0.014)\end{tabular} \\

    Constant
    & 
    \begin{tabular}{@{}c@{}}0.010*** \\ (0.000)\end{tabular}
    &
    \begin{tabular}{@{}c@{}}0.001*** \\ (0.000)\end{tabular}
    &
    \begin{tabular}{@{}c@{}}0.001*** \\ (0.000)\end{tabular}
    &
    \begin{tabular}{@{}c@{}}0.001*** \\ (0.000)\end{tabular} \\

    AIC
    & 
    1141919
    &
    1079141
    &
    1054174
    &
    \textbf{1054109} \\

    \textit{N} & 
    \multicolumn{4}{c}{9,608,320} \\
    
  \bottomrule
  \multicolumn{1}{c}{$* p < 0.05, ** p < 0.01, *** p < 0.001$}
\end{tabular}
\end{table*}

Table \ref{tab:tab9} presents results from logistic regression models predicting users’ relocation status. Once again, the odds ratios of social connectivity are higher than one, leading us to conclude that users connected to big companies are more likely to report recent relocations. Similar to the previous analysis of promotion reports, the best model (lowest AIC) is the one that includes all predictors of user characteristics and the social connectivity $\times$ gender interaction. Once again, we observe a positive interaction effect between gender $\times$ social connectivity. However, as shown in Figure \ref{fig:fig11}, the gender gap in relocation among those that are socially connected to Big Tech firms on LinkedIn is smaller -- indicating a higher payoff to social connectivity for women compared to men. To correct for multiple comparisons, such as multiple variables potentially being statistically significant, all p-values in Table \ref{tab:tab8} and \ref{tab:tab9} have been adjusted using the Bonferroni correction (k = number of variables of each model) \cite{Rice2008}.

\begin{table*}
\centering
  \caption{Estimates (odds ratios) from logistic regression models predicting relocation status by user characteristics.}
  \label{tab:tab9}
  \begin{tabular}{cccccl}
    \toprule
    \begin{tabular}{@{}c@{}}\textit{Dependent variable = Recently relocated} \\ \textit{(ref: Not relocated)}\end{tabular} 
    & 
    \multicolumn{4}{c}{\begin{tabular}{@{}c@{}}\textit{Odds ratio} \\ \textit{(standard error)}\end{tabular}} \\ 
    \midrule
    \begin{tabular}{@{}c@{}}Gender (ref: Male) \\ Female\end{tabular}
    & 
    \begin{tabular}{@{}c@{}}0.787*** \\ (0.006)\end{tabular}
    &
    \begin{tabular}{@{}c@{}}0.787*** \\ (0.006)\end{tabular}
    &
    \begin{tabular}{@{}c@{}}0.838*** \\ (0.005)\end{tabular}
    &
    \begin{tabular}{@{}c@{}}0.795*** \\ (0.006)\end{tabular} \\

    \begin{tabular}{@{}c@{}}Job Seniority (ref: Entry) \\ Senior\end{tabular}
    & - &
    \begin{tabular}{@{}c@{}}1.381*** \\ (0.009)\end{tabular}
    &
    \begin{tabular}{@{}c@{}}1.324*** \\ (0.009)\end{tabular}
    &
    \begin{tabular}{@{}c@{}}1.323*** \\ (0.009)\end{tabular} \\

    Manager
    & - &
    \begin{tabular}{@{}c@{}}1.165*** \\ (0.012)\end{tabular}
    &
    \begin{tabular}{@{}c@{}}1.110*** \\ (0.011)\end{tabular}
    &
    \begin{tabular}{@{}c@{}}1.106*** \\ (0.011)\end{tabular} \\

    Director
    & - &
    \begin{tabular}{@{}c@{}}2.147*** \\ (0.019)\end{tabular}
    &
    \begin{tabular}{@{}c@{}}1.990*** \\ (0.018)\end{tabular}
    &
    \begin{tabular}{@{}c@{}}1.987*** \\ (0.018)\end{tabular} \\

    \begin{tabular}{@{}c@{}}Age Range (ref: 18 to 24) \\ 25 to 34\end{tabular}
    &
    \begin{tabular}{@{}c@{}}1.158*** \\ (0.014)\end{tabular}
    & - &
    \begin{tabular}{@{}c@{}}1.123*** \\ (0.013)\end{tabular}
    &
    \begin{tabular}{@{}c@{}}1.124*** \\ (0.013)\end{tabular} \\

    35 to 54
    &
    \begin{tabular}{@{}c@{}}1.555*** \\ (0.018)\end{tabular}
    & - &
    \begin{tabular}{@{}c@{}}1.381*** \\ (0.016)\end{tabular}
    &
    \begin{tabular}{@{}c@{}}1.383*** \\ (0.016)\end{tabular} \\

     55+
    &
    \begin{tabular}{@{}c@{}}1.461*** \\ (0.022)\end{tabular}
    & - &
    \begin{tabular}{@{}c@{}}1.252*** \\ (0.019)\end{tabular}
    &
    \begin{tabular}{@{}c@{}}1.255*** \\ (0.019)\end{tabular} \\

    \begin{tabular}{@{}c@{}}Social Connectivity (ref: Not connected) \\ Connected to big companies\end{tabular}
    & 
    \begin{tabular}{@{}c@{}}1.247*** \\ (0.009)\end{tabular}
    &
    \begin{tabular}{@{}c@{}}1.191*** \\ (0.009)\end{tabular}
    &
    \begin{tabular}{@{}c@{}}1.210*** \\ (0.007)\end{tabular}
    &
    \begin{tabular}{@{}c@{}}1.150*** \\ (0.008)\end{tabular} \\

    Social Connectivity $\times$ Gender
    & 
    \begin{tabular}{@{}c@{}}1.190*** \\ (0.015)\end{tabular}
    &
    \begin{tabular}{@{}c@{}}1.176*** \\ (0.015)\end{tabular}
    & - &
    \begin{tabular}{@{}c@{}}1.185*** \\ (0.015)\end{tabular} \\

    Constant
    & 
    \begin{tabular}{@{}c@{}}0.010*** \\ (0.000)\end{tabular}
    &
    \begin{tabular}{@{}c@{}}0.011*** \\ (0.000)\end{tabular}
    &
    \begin{tabular}{@{}c@{}}0.009*** \\ (0.000)\end{tabular}
    &
    \begin{tabular}{@{}c@{}}0.009*** \\ (0.000)\end{tabular} \\

    AIC
    & 
    1346296
    &
    1342040
    &
    1340790
    &
    \textbf{1340623} \\

    \textit{N} & 
    \multicolumn{4}{c}{9,626,000} \\
    
  \bottomrule
  \multicolumn{1}{c}{$* p < 0.05, ** p < 0.01, *** p < 0.001$}
\end{tabular}
\end{table*}

\section{Ethical Considerations}
The data collected and used in this paper consist of aggregate and anonymous user counts. As the smallest identifiable unit contains 300 users, we see little to no reidentification risk of individual users. While such aggregate data could, theoretically, still be used to map vulnerable groups, we do not see the general data source -- LinkedIn -- nor the particular targeting attributes posing a danger for this. The data used is also accessible free of charge on LinkedIn's advertising platform through APIs\footnote{See \url{https://worldbank.github.io/connectivity_mapping/linkedin_nbs/interface.html} for a helpful introduction to programmatically collecting such data}. This type of data access can be seen as a  type of Data Collaborative\footnote{\url{https://datacollaboratives.org/}}, providing a non-standard way of enabling partial auditing of large platforms. 

In terms of the privacy \emph{expectation} of users, we believe that this type of data is less problematic than public individual-level data consisting of posts, comments, or pictures. At the same time, we acknowledge that LinkedIn users might be unaware that their data can be accessed and analyzed in this way, even if in anonymous and aggregate form \cite{Anderson2020-tg}.

LinkedIn and all other online advertising platforms we are aware of only support binary female-or-male gender for the ad targeting on the platform, though it leaves some users unclassified. This could be viewed as a form of exclusion, if not erasure, of gender minorities \cite{Bivens2016}. At the same time, this design choice limits the potential use of the advertising platform for targeted harassment of gender minorities. 
 Furthermore, the binary gender used by LinkedIn is automatically inferred, and also draws on pronouns used by other users referring to the user, rather than based on self-identified information. This will undoubtedly result in misclassifications of some users. In a user-facing system, such misclassification has the risk of causing psychological harm by misgendering the user without providing them with an option to self-declare their gender identity \cite{mediumDataViolence, reallifemagCountingCountless}. However, LinkedIn seems to only assign a gender to cases with sufficient confidence. For example, for all the 6,500,000 LinkedIn users in the US working in IT (using the same definition as in the paper), LinkedIn’s advertising platform classifies 2,100,000 as female and 3,900,000 as male, leaving 500,000 as unclassified\footnote{As of Jan 15, 2023, and per \url{https://www.linkedin.com/campaignmanager}.}. For cases that can be mapped with sufficient confidence, openly accessible name-to-gender mappings achieve accuracies of 95-97\% \cite{Santamara2018}. This suggests that for those users for whom a (binary) gender is inferred, the precision is likely to be high. For our population-level study on relative female-vs-male gender gaps, we feel that these data  are able to highlight important, aggregate gender differences that outweigh the harm caused by potential (population-level) gender misidentification. 

\section{Discussion and Conclusion}
Professional networking is important for career progression, yet research has shown that women's offline networks are less advantageous than men's. How these gender differences translate to online spaces, specifically the use of online professional networking platforms, is not well understood. This study examines gender differences in the information technology (IT) industry in two of the largest LinkedIn user populations of the UK and US, leveraging aggregated, anonymised data on the LinkedIn user population from its advertising platform. 
Consistent with previous work using these data \cite{Kashyap2021, verkroost2020tracking, Haranko2018ProfessionalGG, Berte2023MonitoringGG}, we find there are fewer women compared to men on LinkedIn in IT. Female LinkedIn users are younger, less senior, and also less likely to be connected to big companies compared with male LinkedIn users in IT. Yet, they were more likely to report a recent promotion at work. Even in this high-achieving sample, we nonetheless found women were less likely to report a relocation, confirming previous research that highlights women's lower availability to relocate \cite{Baldridge2006, Mansfield2014}.

While the data preclude us from distinguishing whether the observed gender differences in promotion rates reflect differences in propensity to report promotions, or the actual prevalence of promotions among LinkedIn users, we offer two plausible interpretations of these findings, which are not mutually exclusive. First, aligned with prior studies that show positive selection effects by gender on online platforms such as LinkedIn \cite{Kashyap2021, verkroost2020tracking} or Google+ \cite{magno2014international}, women who are on LinkedIn, especially in a highly unequal industry such as IT, may be high-achieving, professionally driven, and positively selected. Second, women on LinkedIn may choose to more actively share recent promotions to their online networks, seeking visibility from the wider professional community at lower costs afforded through online platforms. These benefits from online networking may help women who face greater disadvantage in accessing offline networking due to gendered family or caring responsibilities (e.g., attending conferences or socialising after work), or have smaller offline professional networks. The fact that women often face greater constraints, e.g., related to gendered family expectations, in making job-related decisions is suggested by our finding that even among this sample of professionally motivated women, women are less likely to report relocations. Further suggestive of these constraints is our finding that the social connectivity gap on LinkedIn between men and women is greatest during the childbearing ages. While the lower relocation rate may reflect a lower availability to relocate among women, the differences in promotion versus relocations also suggest the pursuit of different career progression strategies for men and women, which are likely to be shaped by differing choice sets, norms and expectations. 

Our findings add an important gendered nuance to recent research highlighting the value of online professional networking via LinkedIn for job search and mobility processes \cite{rajkumar2022causal, wheeler2022linkedin}. We find that although women not working at Big Tech firms on average have lower social connectivity to those at Big Tech companies than men, the payoffs to online social connectivity for those with these networks are larger for women compared with men. While \citet{rajkumar2022causal} show the causal effects of weak ties on LinkedIn for job search, our findings suggest returns to these ties may be even larger for women, who have conventionally faced greater disadvantages in accessing potentially advantageous network ties in the labour market through traditional forms of networking. We acknowledge nonetheless that the cross-sectional nature of the data implies that our findings cannot be given a causal interpretation, and are susceptible to the potential for reverse causality. Online social connectivity may not be the driver of higher promotion rates, but women who are recently promoted may be seen as more successful and attract more social connections. As such, the formation of social links  reflects a bidirectional process, which makes the interpretation of social connectivity tricky. This process is also likely to be algorithmically mediated, and the acceptance rate of such requests could also correlate with other characteristics, which we are unable to control for in our analyses. Longitudinal data are needed to better disentangle these mechanisms underlying social connectivity and job progression.

We acknowledge that the data come with additional limitations. The audience counts obtained from the marketing platform may include fake accounts, and also include misrepresented or inaccurate affiliations. Moreover, there is a sparsity-related limitation as LinkedIn’s advertising platform does not provide counts below 300. Currently, our data has 36 out of 192 (19\%) sparse values. If we were to disaggregate by the two countries (US vs.\ UK), then we would have 116 out of 384 (30\%) sparse values. As we felt that this level of sparsity with values missing not at random would be too high, we decided to combine the two countries. Further, many of the targeting categories for which counts are provided are algorithmically inferred, and are vulnerable to biases. For example, people with non-standard careers who start university later in life, might be misclassified as being younger than they are. Greater transparency and documentation from platforms about the data-generating process and algorithms underlying these data can be helpful to better understand and address these biases. Even within our existing (binary) analysis of gender, we recognise that men and women are not homogeneous groups, and acknowledge that much heterogeneity exists in how workplace structures may differentially affect the experiences and professional trajectories of individuals -- e.g., by race, immigration status, or sexual orientation, and how these intersect. 

Nonetheless, our findings expand on previous research about gender gaps on LinkedIn, by exploring the additional dimensions of social connectivity and how these are associated with job progression behaviours such as recent promotions and relocations. They contribute to a growing body of work showing the potential value of online professional networks for employment behaviours, but highlight the need to integrate a gender perspective to understand the differential impacts of online platforms on social and economic domains. With the growing digitalisation of work, and also increasing levels of remote work \cite{Remote2021}, online networking is likely to become even more central to job progression and mobility processes. This increasing use of online networking may help to mitigate gender gaps in the labour market. In turn, policies that integrate online professional networking within educational and job training programmes have the potential to help benefit disadvantaged groups. Moreover, for employers, seeking potential candidates through online platforms may serve to bring a broader pool of candidates to their attention than those through traditional network-based contexts, e.g. through conferences or events. For researchers, our study motivates further studies of user behaviours on LinkedIn in its own right as the largest professional networking platform, but also studies that examine how online networking is experienced and used by different disadvantaged social groups, and whether they reproduce or alter social inequalities experienced by them. 

\section{Competing Interests}
The authors declare that they have no competing interests.


\bibliography{aaai22}

\end{document}